\documentclass[preprint,secnumarabic,amssymb, nobibnotes, aps, prb]{revtex4}
\usepackage{graphicx}

\begin{document}

\title{ Microwave excitations associated with a wavy angular dependence of the spin transfer torque : model and experiments}%

\author{O. Boulle}
\altaffiliation{Present adress: Universit\"at Konstanz, Universit\"atsstr. 10, D - 78457 Konstanz, Germany : olivier.boulle@uni-konstanz.de}
\affiliation{Unit\'e mixte de physique CNRS/Thales and Universit\'e Paris Sud-11, Route d\'epartementale  128, 91767  Palaiseau, France}

\author{V. Cros}
\altaffiliation{Corresponding author : vincent.cros@thalesgroup.com}
\affiliation{Unit\'e mixte de physique CNRS/Thales and Universit\'e Paris Sud-11, Route d\'epartementale  128, 91767  Palaiseau, France}

\author{J. Grollier}
\affiliation{Unit\'e mixte de physique CNRS/Thales and Universit\'e Paris Sud-11, Route d\'epartementale  128, 91767  Palaiseau, France}
\author{L.G. Pereira}
\altaffiliation{Present adress: Instituto de Fisica, UFRGS, 91501-970 Porto Alegre, RS, Brazil} 
\affiliation{Unit\'e mixte de physique CNRS/Thales and Universit\'e Paris Sud-11, Route d\'epartementale  128, 91767  Palaiseau, France}

\author{C. Deranlot}
\affiliation{Unit\'e mixte de physique CNRS/Thales and Universit\'e Paris Sud-11, Route d\'epartementale  128, 91767  Palaiseau, France}
\author{F. Petroff}
\affiliation{Unit\'e mixte de physique CNRS/Thales and Universit\'e Paris Sud-11, Route d\'epartementale  128, 91767  Palaiseau, France}
\author{G. Faini}
\affiliation{Phynano team, Laboratoire de Photonique et de Nanostructures LPN-CNRS,
Route de Nozay, 91460 Marcoussis, France}
\author{J. Barna\'s}
\affiliation{Department of Physics, Adam Mickiewicz University, Umultowska 85, 61-614 Poznan, Poland}
\author{A. Fert} 
\affiliation{Unit\'e mixte de physique CNRS/Thales and Universit\'e Paris Sud-11, Route d\'epartementale  128, 91767  Palaiseau, France}

\date{\today}%
\begin{abstract}
The spin transfer torque (STT) can lead to steady precession of magnetization without any external applied field  in magnetic spin valve where the magnetic layer have very different spin diffusion length. This effect is associated with an unusual angular dependence of the STT, called "wavy" (WAD-STT), predicted in the frame of diffusive models of spin transfer. In this article, we present a  complete experimental characterization of the magnetization dynamics in the presence of a WAD-STT. The results are compared to the prediction of the magnetization dynamics obtained by single domain magnetic simulations (macrospin approximation). The macrospin simulations well reproduced the main static and dynamical experimental features (phase diagram, R(I) curves, dependence of frequency with current and field)  and suggest that the dynamical excitations observed experimentally are associated with a large angle out-of-plane precession mode. The present work validates the  diffusive models of the spin transfer and underlines the role of the spin accumulation and the spin relaxation effects on the STT.

\end{abstract}

\maketitle
\section{Introduction}

A spin polarized current can exert a torque on the magnetization of a ferromagnetic body, via transfer of spin angular momentum. This spin transfer effect, originally introduced  by J.Slonczewski and L.Berger~\cite{Slonczewski96JoMaMM,Berger96PRB}, has opened a new route to manipulate a magnetization using an electrical current. Today, it is the subject of extensive experimental and theoretical research motivated not only by its fundamental interest but also by promising applications in the field of magnetic memories and microwave devices for future telecommunication systems. In most experiments, this effect has been   studied in pillar shaped F1/NM/F2 trilayers in which  a magnetic layer F1 with a fixed magnetization is used to prepare a spin polarized current injected in a magnetically free  magnetic layer F2. When the spin polarized current enters the free layer, the part of the spin current that is non-collinear to the magnetization of the free layer is absorbed and transferred to the local magnetization; this spin transfer is equivalent to a torque exerted on magnetization. In \textit{standard} structures such as Co/Cu/Co or NiFe/Cu/NiFe,  at zero or low applied field, the spin transfer torque (STT) leads to  an irreversible switching of the magnetization from one static configuration to another one~\cite{Katine00PRL,Grollier01APL} (so called CIMS effect). 
 For larger  fields (typically higher than the coercive field of the free layer), the spin transfer torque compensates for the damping torque resulting in a steady precession of magnetization around the local internal field. Owing to the giant magnetoresistance (GMR) effect, this high frequency magnetization precession goes with voltage oscillations in the microwave range~\cite{Kiselev03N,Rippard04PRL}.
These two behaviors (magnetization switching and precession) open two branches of potential technological applications of the spin transfer effect: first, a new and more reliable way to write the magnetic bit in magnetic memories and second, a new type of sub-micrometric microwave oscillators for future telecommunication devices combining high quality factor and large frequency agility with current and field ~\cite{Rippard04PRL}.

Following the original approach of Slonczewski~\cite{Slonczewski96JoMaMM}, several theoretical studies  underlined that the STT is directly related to the spin polarization of the spin current entering the magnetic free layer and more precisely to the absorption of the spin current's component  transverse  to the magnetization~\cite{Stiles02PRB,Brataas06CM}. In the first generation of spin transfer models~\cite{Slonczewski96JoMaMM,Berger96PRB,Waintal00PRB}, this spin polarization was calculated in a ballistic assumption. The magnetic multilayer is contacted to reservoirs with uniform chemical potentials and the spin polarization of the current arises from the spin dependent reflections at the interfaces of the multilayer and from spin dependent scattering inside the magnetic layers. However, since the study of the CPP-GMR~\cite{Valet93PRB,Yang94PRL}, it is well known that, in this geometry, the spin polarization of the current is strongly affected by the spin accumulation which is controlled by spin relaxation effects in the whole structure. To correctly describe the electronic transport, diffusive models that also take into account this relaxation must be considered~\cite{Valet93PRB}. Indeed recent spin transfer models do take into account these diffusive transport aspects~\cite{Stiles02PRB,Fert04JoMaMM,Barnas05PRB,Brataas06CM,Slonczewski02JoMaMM,Manchon06PRBa}.

\begin{figure}[p]
	\centering
		\includegraphics[width=0.5\textwidth]{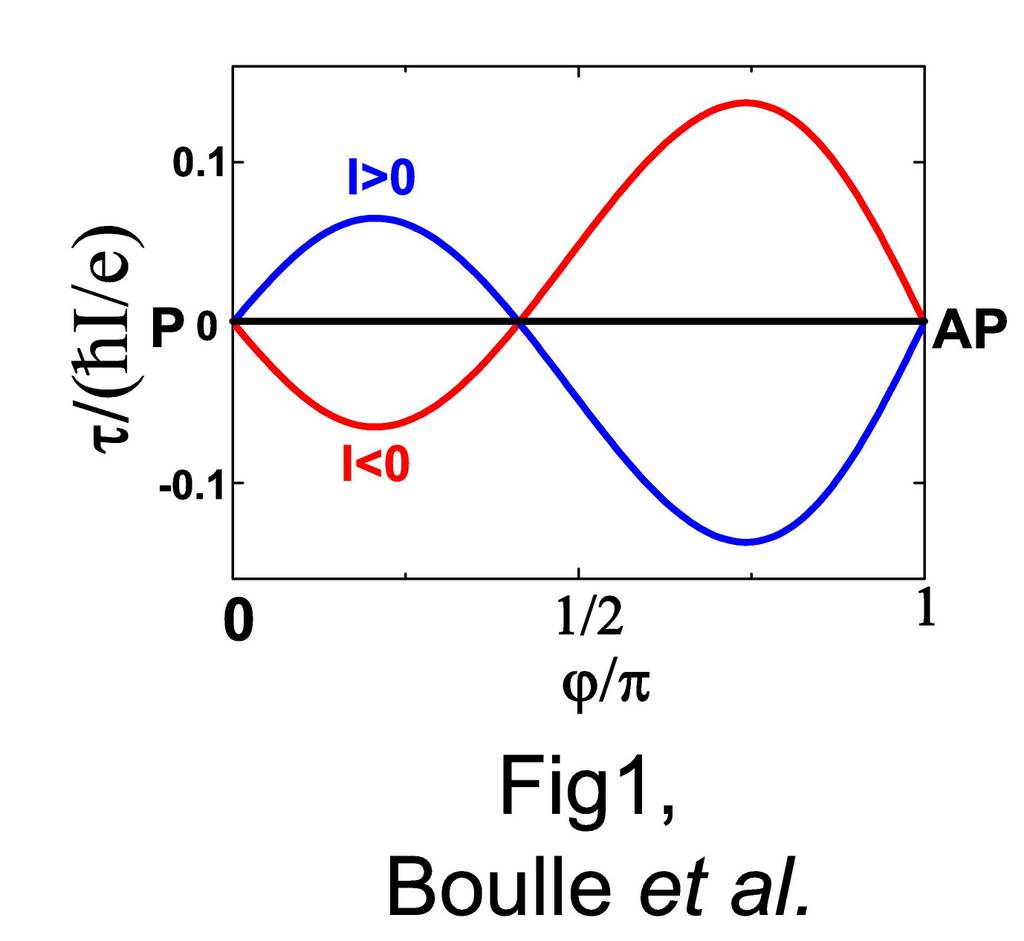}
	\caption{Magnitude of the spin transfer torque $\tau$ on the free Py layer of a Au(infinite)/Cu(5 nm)/Py(8 nm)/Cu(10 nm)/Co(8 nm)/Ta(10 nm)/Cu(infinite) multilayer as a function of the angle $\varphi$ between the magnetizations of the free Py and fixed Co layers, calculated with the Barna\'s-Fert model~\cite{Barnas05PRB}\,($\tau(\varphi)=-P(\varphi)\sin\varphi I\hbar/(2e)$). Electrons flow from the Co layer to the  Py layer for a positive current $I$. }
	\label{fig0}
\end{figure}

The first experimental evidence of diffusive effects on the STT was given by Urazhdin \textit{et al.}~\cite{Urazhdin04APL} and Alhajdarwish \textit{et al.}~\cite{AlHajDarwish04PRL}. They demonstrated the influence of the \textit{spin dependent scattering} by magnetic impurities on  the STT features in the CIMS regime at low magnetic field : by doping the polarizing layer with impurities, the amplitude and even the sign of the torque was changed. More recently, we demonstrated experimentally the influence of the \textit{spin relaxation effect} on the STT~\cite{Boulle07NP}. We showed that by acting on the distribution of spin relaxation in the structure through different spin diffusion lengths in the two magnetic layers, the STT and thus the dynamics of magnetization induced by this torque can be strongly modified compared to \textit{standard} structures.
This was made by studying the magnetization dynamics induced by the STT in a Co(8~nm)/Cu/Py(8~nm) (Py=Ni$_{80}$Fe$_{20}$) nanopillar through both static and high frequency magneto-transport measurements. In such a structure, an unusual dependence of the STT with the angle $\varphi$ between both magnetization is predicted in the frame of diffusive models of spin transfer~\cite{Barnas05PRB,Barnas06MSaEB,Gmitra06PRL} (see Fig.~\ref{fig0}). This so called Wavy Angular Dependence of the Spin Transfer Torque (WAD-STT) arises from the modification of the spin accumulation profile in the structure caused by the different ratios \textit{thickness /spin diffusion length } of the magnetic layers.
It is characterized by a change of sign of the STT between 0 (parallel configuration ($P$)) and $\pi$ (antiparallel  configuration ($AP$)). This angular dependence modifies the stability of the static states $P$ and $AP$: for one polarity of the current (negative current in our convention), both states are stabilized by the STT whereas they are destabilized for the opposite polarity (positive current). In the latter case, it is predicted that the magnetization precesses even at zero external magnetic field. These predictions  were validated experimentally by the present authors by measuring the microwave emission caused by the steady precession of magnetization induced by the STT at zero (or low) applied field~\cite{Boulle07NP}.

%keepaspectratio=1

 In this manuscript, we present a complete characterization and analysis of the magnetization dynamics in the presence of a WAD-STT. The results are compared to the prediction of the magnetization dynamics obtained by single domain magnetic simulations (macrospin approximation). We show that the main static and dynamical experimental features such as current-field phase diagram, variation of resistance with current, variation of frequency with current and field can be reproduced by simulation, at least qualitatively and for some features quantitatively. Therefore, the present work validates the  diffusive models of the spin transfer~\cite{Stiles02PRB,Fert04JoMaMM, Zhang04PRL,Brataas06CM}. 

The paper is organized as follows. We first described the predictions of the macrospin simulations taking into account the WAD-STT. In a second section, we present the results of low field static and high frequency measurements obtained on a Co(8~nm)/Cu/Py(8~nm) nanopillar in which a WAD-STT is predicted. These results are compared with the predictions of the macrospin simulations and discussed in a third part . 

\section{Macrospin simulations with a wavy angular dependance of the spin transfer torque (WAD-STT)}
\subsection{Method}
In this section, we present the results of the simulations for the magnetization dynamics of the Py  free  layer of a Co(8~nm)/Cu/Py(8~nm) trilayer, on the basis of the predicted \textit{wavy} angular dependence of the torque (labelled as WAD-STT). 
These calculations have been carried out in the macrospin approximation, \textit{i.e} by assuming the magnetization to be uniform during its motion. 
 The magnetization of the Py free layer is described by a unit vector $\vec m=m_x\vec u_x+ m_y \vec u_y+ m_z\vec u_z$ and makes an angle $\varphi$ with  the magnetization of the  Co layer which is supposed fixed and aligned along the x axis. As in experiments, the Py free layer has got an elliptical  shape of dimensions 100x155 nm$^2$ and a thickness $t_{Py}$ of 8 nm.  This results in an in-plane uniaxial  anisotropy  field $\vec H_{an}=H_{an} m_x \vec u_x$ aligned along the direction of the long axis $\vec u_x$ of the ellipse and a demagnetizing field $\vec H_d=-H_dm_z \vec u_z$, with  $\vec u_z$ perpendicular to the plane of the layer. The external field $\vec H_{app}=H_{app} \vec u_x$ is applied in the plane of the layer and  aligned along the direction $\vec u_x$ of the uniaxial anisotropy field.
The time dependent trajectory  $\vec m(t)$  is found by solving the modified Landau-Lifschitz-Gilbert (LLG) equation including the STT $\vec{\tau}$:
\begin{equation}
\label{eq:LLG}
\frac{d\vec{m}}{dt}=-\gamma_0\,\vec{m}\times\vec{H}_{f}+\alpha\,\vec{m}\times\frac{d\vec{m}}{dt}-\frac{\gamma_0}{\mu_0M_sV}\vec\tau
\end{equation}

 Here $\gamma_0$ is the absolute value of the gyromagnetic ratio, $\alpha$ the parameter of the Gilbert damping, M$_s$ the saturation magnetization, V the volume of the Py nanomagnet and $\vec H_{f}$ an effective field which includes $\vec H_{app}$, $\vec H_{an}$ and $\vec H_{d}$. 
In order to characterize the influence of thermal activation on the magnetization dynamics, the simulations were carried out both at T=0~K and T=300~K. Thermal effects are simulated by  introducing a randomly fluctuating field $\vec H_T$~\cite{Brown63PR,Russek05PRB} with $\mu_0\vec H_T=\sqrt{2k_BT\alpha/VM_s\gamma\Delta t}\vec H_{al}$, where $\vec H_{al}$ is a random gaussian  field with $<\vec H_{al}>=0$ and $<H_{al}^2>=1$,  $k_B$ is the Boltzmann's constant and $\Delta t$ the integration step. The magnetic parameters used for the simulations are described in appendix~\ref{appendix1} with computational details.
 The spin transfer term is defined as $\vec \tau=P(\varphi)\frac{I\hbar}{2 e}\,\vec{m}\times\vec{m}\times\vec{u_x}$, $I$ is the current   (defined as positive when the electrons flow from the fixed Co layer to the Py free layer), $e$ is the absolute value of the electron charge. The polarization factor $P(\varphi)$ is derived from the angular dependence of the torque calculated from the Barna\'s-Fert model (see Fig.~\ref{fig0}).
    
The displayed phase diagrams are constructed by calculating the mean values over the integration time of $\sigma_{mx}=\sqrt{<m_x^2-<m_x>^2>}$ after relaxation  ($\sigma_{mx}$ gives an evaluation of the amplitude of precession). These two quantities allow to determine the borders between the different static and dynamical states. All presented simulations have been carried out for a positive current. No excitations are predicted by the simulations for a negative current\footnote{According to our notation, the electrons are going from the Co fixed layer to the NiFe free layer for a positive current. Note that the excitations of the parallel state by a positive current is in contrary with the standard situation in which a negative current excites the parallel state.}.

\subsection{Results}
\subsubsection{T=0~K}

\begin{figure}[p]
	\centering
		\includegraphics[width=1\textwidth]{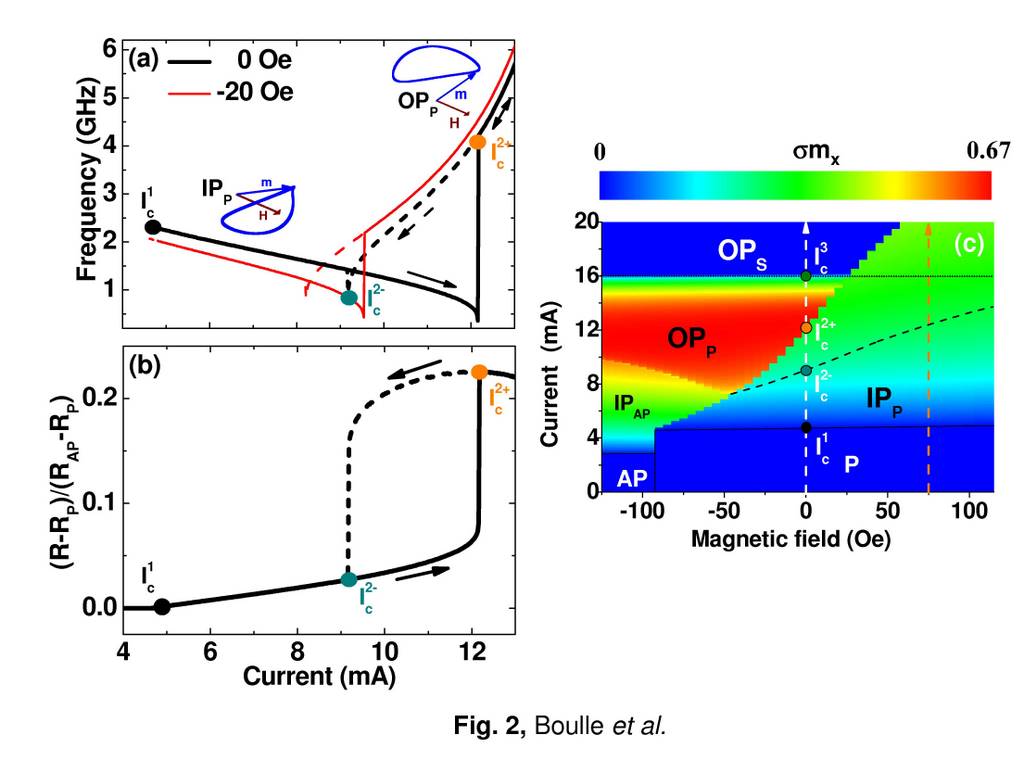}
	\caption{T=0~K. (a) Frequency of precession as a function of current for several applied field for increasing (plain line) and decreasing (dashed line) current.  (b) Normalized resistance as function of current for H=0 for increasing (plain line) and decreasing (dashed line) current. (c)  Dynamical phase diagram  for increasing current ($\sigma_{mx}$  in color scale) and decreasing current (dashed and dotted frontiers). 	For decreasing current, the dotted white line corresponds to the frontier of the $OP_S$ with the $OP_P$ state  and the dashed black line to the frontier of the $OP_P$ state with the $IP_P$ state.  Plain black lines define the frontiers of the P and AP state. Static and dynamical states are defined in the text.}
	\label{fig1}
\end{figure}

In Fig.~\ref{fig1}(a) and (b), we show respectively the variation of the frequency and the resistance as a function of the current for zero applied field (black curves). 
 Starting from an initial $P$ configuration and increasing current (plain line), the magnetization goes into sustained precession around the in plane effective field (in plane precession mode $IP_P$) for a  threshold  current $I_c^1$ (black  point at I=4.8~mA) with a precession frequency (f=2.28~GHz) close to the  Kittel's frequency of  the FMR small angles precession (f=2.23~GHz).  
 When the current is increased above $I_c^1$, the angle of precession increases but the frequency decreases ("red shift regime"). This behavior, i.e. the decrease of frequency
with increasing current is a nonlinear effect due to the dependence of the frequency on the precession amplitude~\cite{Slavin05ITOM,Rezende05PRL,Bertotti05PRL}. In this mode, only a small increase of the resistance with increasing current is predicted,  although a large angle precession can be reached. For a threshold current $I_c^{2+}$  (orange point at I=12~mA), the magnetization dynamics changes abruptly to an out-of-plane precession mode (labelled as $OP_P$ ). This transition is associated with a jump in frequency and a large increase of the resistance. In this $OP_P$ regime, the magnetization precesses around the demagnetizing field and the frequency is set by the mean out-of-plane component of magnetization $<m_z>$. For increasing current,  the trajectory  goes away from the plane of the layer  (higher $<m_z>$) and the frequency increases ("blue shift regime"). For higher current (I$>13$~mA, not shown), the magnetization precesses faster (up to 20~GHz) around circular trajectories of decreasing radius while current is increased. These trajectories converge for even higher current to a quasi-static out-of-plane state $OP_S$.  As shown in Fig.~\ref{fig1}(a-b), the transition between the $IP_P$ and the $OP_P$ precession modes is irreversible): the critical current for this transition is higher for increasing current ($I_c^{2+}$, orange point and plain line)  than for decreasing current ($I_c^{2-}$, blue point and dashed line), which results in  hysteretic R(I) and f(I)  curves.

In addition, we display in Fig.~\ref{fig1}(a) the variation of the frequency as a function of the current for a higher negative applied field H = -20~Oe (red curve). In the $IP_P$ mode, the black curve (H=0~Oe) is above the red curve (H=-20~Oe): the frequency increases with field.  On the contrary, in the $OP_P$ mode, the black curve is below the red curve and the frequency decreases with field:  the in-plane field pulls magnetization in the plane of the layer, decreasing  $<m_z>$ and therefore the frequency.

We present in  Fig.~\ref{fig1}(c) a calculated current-field dynamical phase diagram for increasing current starting from an initial $P$ configuration. The amplitude of precession ($\sigma_{mx}$)  is plotted in a color scale. Several regimes can be defined.
First, at high positive field ($H>+25$~Oe) (see the orange dashed  line in Fig.~\ref{fig1}(c)), only the $IP_P$ mode is observed for  $I_c^1<I< I_c^3$.  For $I> I_c^3$, the magnetization goes into the quasi-static out-of-plane precession  state $OP_S$. In a second regime at low field i.e. for fields smaller than the coercive field (see the white dotted line on Fig.~\ref{fig1}(c)), the out-of-plane precession mode $OP_P$ appears for I above $I_c^{2+}$  with $I_c^1<I_c^{2+}<I< I_c^3$. The amplitude of precession (red color) and therefore the emitted power is maximum in this mode.  The aforementionned hysteretic behavior  appears also on the phase diagram: the $OP_P$/$IP_P$ frontier  is not the same for \emph{increasing} current (colored frontier, $I_c^{2+}$) and \emph{decreasing current}  (dashed lines, $I_c^{2-}$). At higher current and field ($H>+25$~Oe), the transition  to the $OP_S$ state is also irreversible: for increasing current (colored frontiers), magnetization switches from IP$_P$  to $OP_S$  and for decreasing current from $OP_S$  to $OP_P$  (dotted lines).

Finally, for even a lower applied field (H$<-50$~Oe), before reaching the $OP_P$ mode, the magnetization first precesses in-plane around the $AP$ state  (mode denoted as $IP_{AP}$). Note that in this mode, the frequency increases with current and decreases with (positive) field. Starting from an initial $AP$ state (H$<-90$~Oe), this precession state $IP_{AP}$ is the first observed when current is increased from zero.

To summarize, the  magnetization dynamics in a WAD-STT structure is very different from what is predicted for a \textit{standard} angular dependence structure: 
\begin{itemize}
	\item For $H>H_c$, where $H_c$ is the coercive field of the free layer, the precession modes are predicted for  a positive current for a WAD-STT structure in our current convention, whereas they are obtained for a negative current for a \textit{standard} structure. 
		\item For $H<H_c$, in the \textit{standard} case, only small angle $IP_P$ precessions  around the in-plane effective field are predicted in a very narrow range of current before the magnetization switches to the other stable static state ($P$ or $AP$) due to the increase of the precession amplitude with current~\cite{Sun00PRB, Stiles05}. In the case of a WAD-STT, both static states ($P$ or $AP$) are unstable and this in-plane-precession mode is present on a much larger range of current. In a similar manner to what is observed for a standard angular dependence at high field, large angle in-plane trajectories bifurcate into  out-of-plane trajectories at higher current~\cite{Bertotti05PRL,Stiles05}. One can note that this large angle out-of-plane precession is predicted even in the absence of any external  field. It is therefore possible in this structure to emit microwaves at zero field, only by injecting a current in the structure. 
\end{itemize}

\subsubsection{T=300~K}

\begin{figure}[p]
	\centering
		\includegraphics[width=1\textwidth]{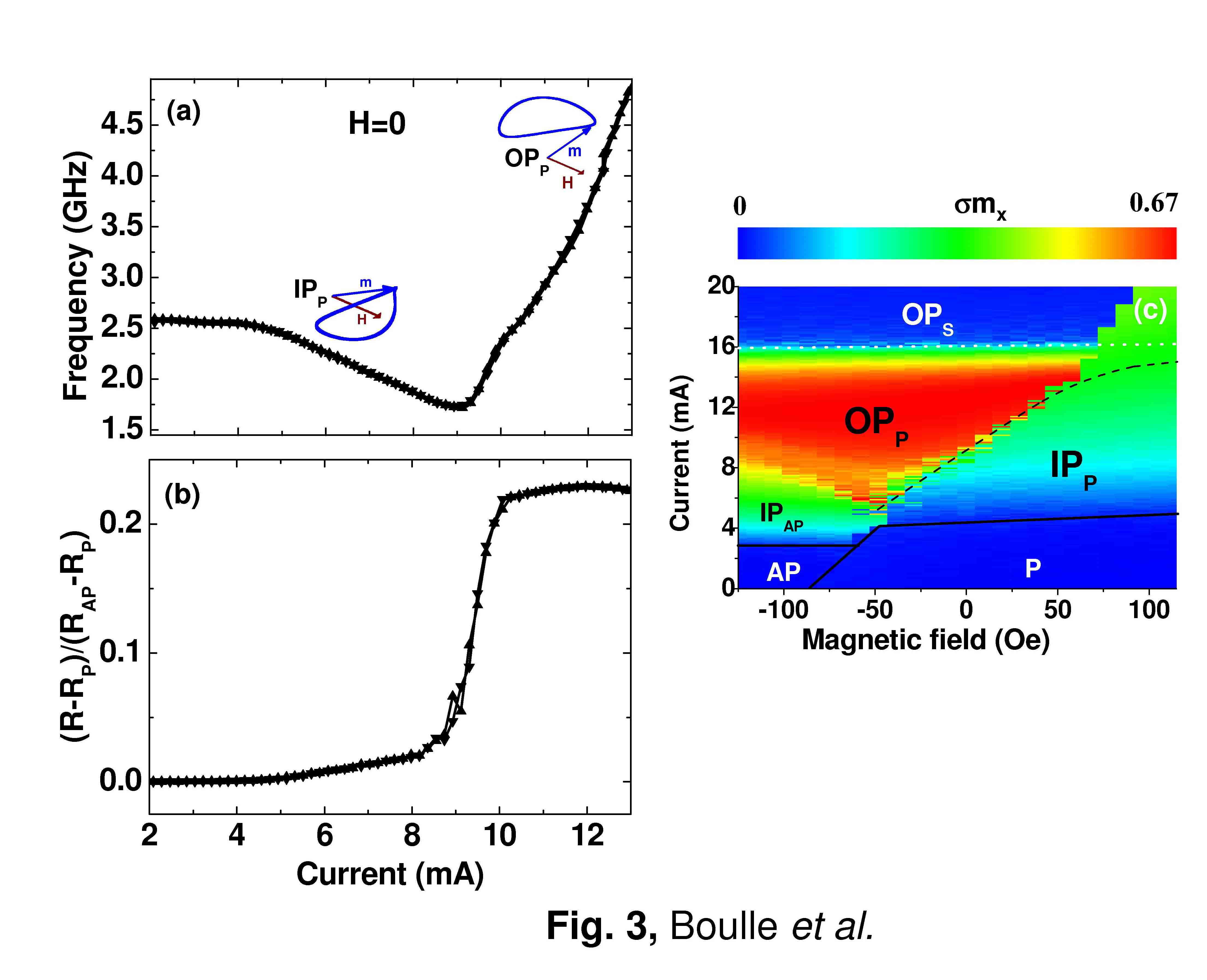}
	\caption{  T=300~K. (a-b) Frequency of precession (a) and normalized resistance (b) as a function of current for increasing (upward triangles) and decreasing (downward triangles) current for H=0. (c)  Dynamical phase diagram  for increasing current ($\sigma_{mx}$  in color scale) and decreasing current (dashed and dotted frontiers).}
	\label{fig2}
\end{figure}

The main effect of temperature is the almost complete disappearance of the irreversibility associated with the  transition from the $IP_P$ to the $OP_P$ precession mode. The thermal energy allows the magnetization to go across the energy barrier that separates these two states.  
 This appears on the phase diagram of Fig.~\ref{fig2}(c): the frontiers between the two states are quasi-identical for increasing (colored frontier) or  decreasing current (dashed lines), except at large applied field ($H>60$~Oe). As a consequence, this reversible behavior leads to an expansion of the current-field region in which the $OP_P$ precession mode occurs: for a given value of the applied field, the critical current for the transition from $IP_P$ to $OP_P$  is lower at T=300~K\,\footnote{We check this decrease is not due to the different value of H$_{an}$ used for T=0~K and T=300~K}. 
The reversibility of this transition results in non-hysteretic f(I) and R(I) curves as shown in Fig.~\ref{fig2}(a-b))\,\footnote{At low current below $I_c^1$, the plotted frequency corresponds to small  elliptical precession of magnetization induced by thermal fluctuations  whose frequency (2.58~GHz) is equal to the Kittel FMR frequency (2.58~GHz). The small variation of the frequency observed at low current for the in-plane precession mode between 0 and 300~K is due to the higher anisotropy field used in the simulation for T=300~K (H$_{an}=120$~Oe instead of 90~Oe.). }. The effect of thermal energy appears in particular in the transition zone between the two states and especially for currents close to the minimum in frequency (I$\approx9.1$~mA). In this zone, the thermal energy makes the system unstable and telegraph noise with a characteristic time of about 10~ns is observed between the state $IP_P$ and the two degenerated  out-of-plane trajectories which are symetrical around the (xy) plane. 
To conclude this section, the temperature does not modify the main characteristics of the dynamics in a WAD-STT structure, such as a steady precession at zero applied field.

\section{Experimental results and comparison with macrospin simulations}  

 We present a detailed experimental characterization of the magnetization dynamics induced by STT on a Co(8~nm)/Cu(10~nm)/Py(8~nm) nanopillar in which a WAD-STT  is predicted. This dynamics was studied by measurements of the high frequency voltage oscillations generated by the magnetization precession with a spectrum analyzer after amplification. To directly compare the experimental results with simulations, we consider for experiments  an effective external field $H_{eff}$  that 
takes into account the contribution from the  dipolar field exerted on the Py free layer ($H_{dip}\approx-43$~Oe).

\subsection{Initial configuration : $P$}

\begin{figure}[p]
	\centering
		\includegraphics[width=0.8\textwidth]{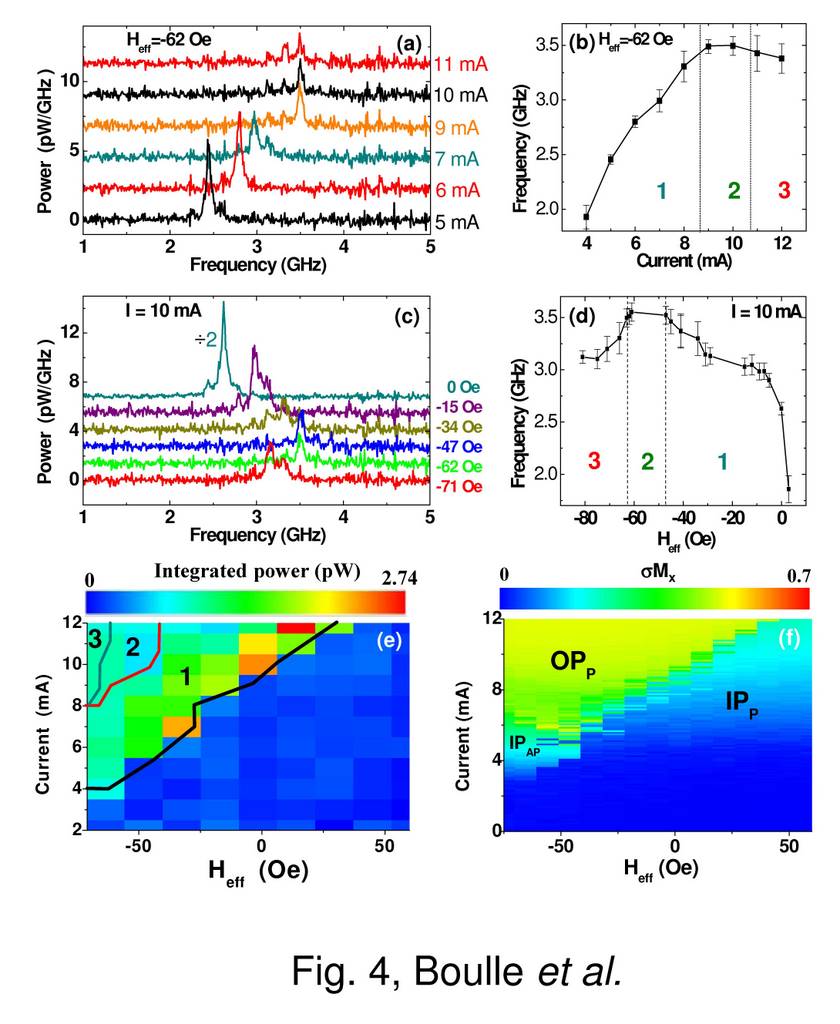}
	\caption { (a): Microwave power spectra for several values of injected current and H$_{eff}=-63$~Oe. (b) Frequency associated to the peaks in the microwave spectra as a function of current for H$_{eff}=-63$~Oe. (c): Microwave power spectra for several values of H$_{eff}$ and I=10~mA. (d) Frequency associated to the peaks in the microwave spectra as a function of the H$_{eff}$ for I=10~mA. H$_{eff}$ is defined as H$_{eff}=H_{app}+H_{dip}$ with $H_{dip}=-43$~Oe. In (b) and (d), error bars correspond to the line width of the peak. (e) Experimental dynamical  current-field phase diagram for increasing current starting from an initial $P$ state. The microwave power is plotted in color scale. The dynamical regimes denoted as 1, 2 and 3 are defined in the text. (f) Simulated current-field phase diagram for increasing current starting from an initial $P$ state. The amplitude of oscillation ($\sigma_{mx}$) is plotted in color scale.}
	\label{fig3}
\end{figure}

In Fig.~\ref{fig3}(a), we display the microwave spectra measured for several values of injected current and a fixed effective applied field $H_{eff}= -62$~Oe . The corresponding variation of the frequency with current at this field is plotted in Fig.~\ref{fig3}(b). The analysis of the frequency dependence with current allows the definition of three different dynamical regimes. At low current (4~mA$\leqslant I \leqslant9$~mA), we observe that the frequency increases with current i.e. a blue shift with current (regime 1). Then it stays approximately constant for 9~mA$\leqslant I \leqslant10$~mA) (regime 2). At higher current ($I>10$~mA), a slight decrease of frequency with current is observed, i.e. a red shift with current (regime 3)\,\footnote{This decrease appears more clearly at even lower field ($H_{eff}\leq-75$~Oe) close to the switching field (not shown).}. Three corresponding behaviors are observed in the frequency dependence with field as illustrated in Fig.~\ref{fig3}(c) by the microwave spectra measured for a constant current of 10~mA and several values of applied field. The corresponding frequency are plotted Fig.~\ref{fig3}(d)). In the low field range (-47~Oe$\leqslant H_{eff}\leqslant2$~Oe), the frequency decreases with field (regime~1). At higher negative field (-62~Oe$\leqslant H_{eff} \leqslant-47$~Oe), the frequency remains approximately constant (f$\approx3.5$~GHz) with field  (regime~2). Then in a narrow experimental window for $H_{eff}$  very close to the switching field, the frequency decreases with field (regime~3).

In Fig.~\ref{fig3}(e), we present  a current-field dynamical phase  with the microwave power plotted in color scale. As predicted by the model, the microwave power is emitted for a \emph {positive} current  and at low field ($H_{eff}<30$~Oe), i.e smaller than the coercive field, and in particular at \emph{zero effective field}. This behaviour is very different from the one typically  observed in standard structures, where steady precession associated with microwave emission is observed at \emph{high field}, i.e higher than the coercive field, and for a  \emph{ negative} current~\cite{Kiselev03N, Kiselev05PRB}.
In addition, we plot on the diagram in plain lines the frontiers between the different dynamical regimes presented above\,\footnote{These frontiers were constructed by considering  the dependence of the emission frequency with current at fixed field from more than a hundred microwave spectra.  Similar frontiers are obtained if the phase diagram is constructed from the dependence of the emission frequency with field at fixed current (not shown).}.  In regime 1, the frequency increases with current and decreases with field. It is present in the largest part of the region of the phase diagram where a power emission is observed. This behaviour is indeed the only one observed at low current (from I$\geqslant4$~mA) and close to zero effective field. The regime 2 appears at low applied field and for higher currents. It is characterized by a frequency that remains approximately constant with the current and also with the applied field. Finally, the regime 3 in which the frequency slightly decreases with current (and increases with field) is observed in the top-left corner of the  diagram corresponding to large current and applied fields close to the switching field ($H_{eff}\approx-90$~Oe).

We can directly compare the experimental results to the simulated dynamical phase diagram shown in Fig.~\ref{fig3}(f). The current-field region in which an out-of-plane precession $OP_P$ mode is predicted (red color in Fig.~\ref{fig3}(e)) coincides with the zone in which the high frequency power is measured (dynamical regime 1, 2 and 3. Note that the field dependence of the experimental critical currents is in excellent agreement with the one associated with the $IP_P$/$OP_P$ transition in the simulations. In addition no microwave power was measured in  region where an $IP_P$ mode is predicted . We believe that, most probably, the weak microwave emission associated with these $IP_P$ mode in a WAD-STT structure is too small to be measured with our experimental setup (this point will be discussed in the section~\ref{sec:discussion}). Finally,  the predicted $IP_{AP}$ precession mode  in the lower left corner of the diagram characterized by  a red shift regime with current and moderate precession amplitude was not observed experimentally.

\subsection{Initial  configuration : $AP$}

\begin{figure}[p]
	\centering
		\includegraphics[width=1\textwidth]{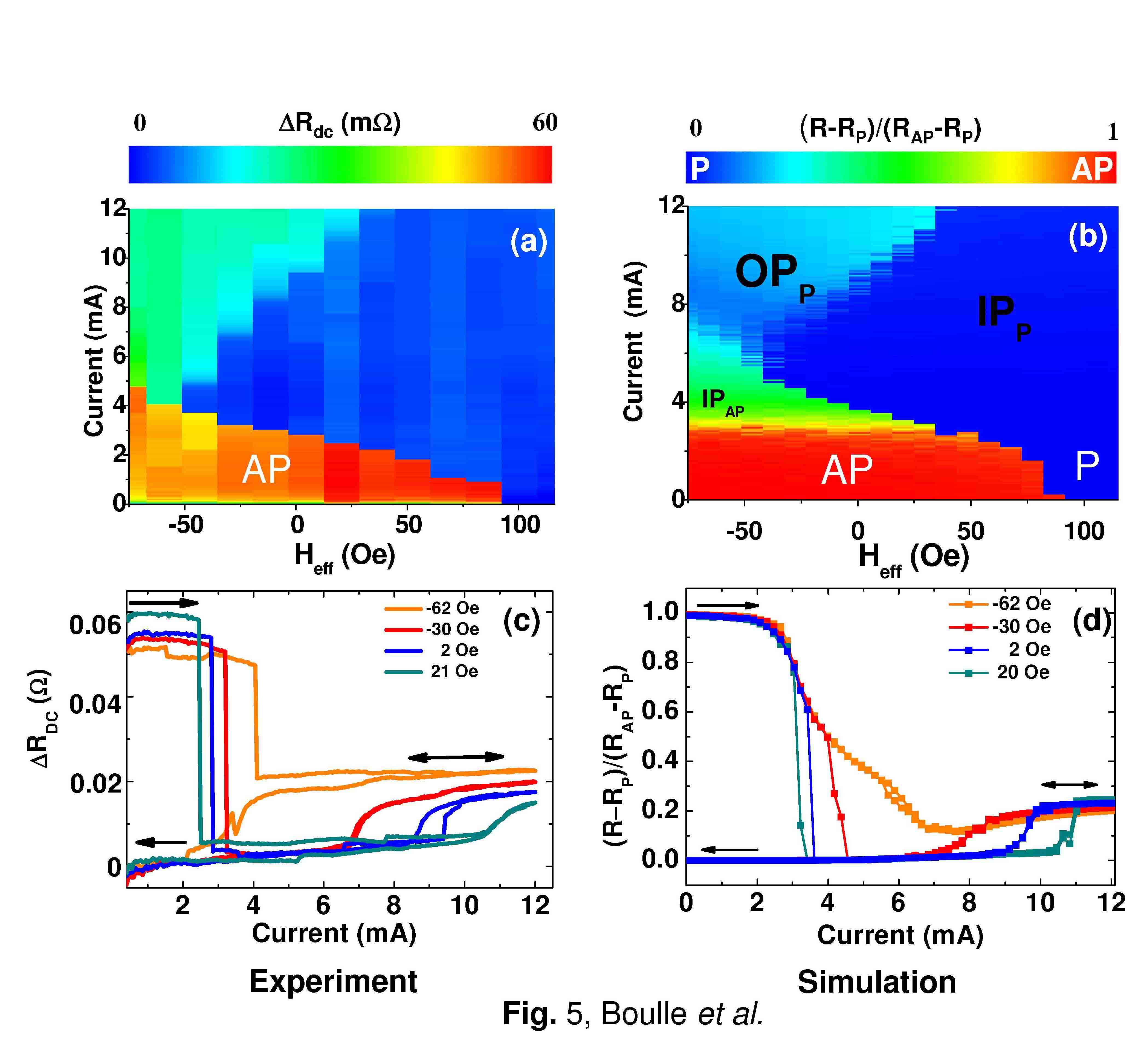}
	\caption{ (a) Experimental  phase diagram for increasing current starting from an initial $AP$ state. The normalized resistance of the sample $\Delta R$ is plotted in color scale. (b) Corresponding simulated phase diagram (normalized resistance in color scale). (c): Measured resistance as a function of current for several value of $H_{eff}$ starting from an initial AP state for successively increasing and decreasing current. (d): Corresponding simulated normalized resistance  vs current curves. The experimental normalized resistance $\Delta R$  is obtained by substracting from the experimental $R$ versus $I$ curves a reference curve to remove the changes in resistance due to Joule heating. }
	\label{fig4}
\end{figure}

We can go deeper into the compared analysis of experiments and simulations by looking at some additional features observed when one starts from the antiparallel $AP$ configuration. In Fig.~\ref{fig4}(a) and (b), we show respectively the experimental and  the simulated phase diagrams for increasing current starting from \emph{an initial \emph{AP} configuration}. The experimental diagram was constructed by measuring R(I) curves for different values of applied field and increasing the current starting from the high resistance $AP$ configuration. The resistance level is plotted in colour scale in Fig.~\ref{fig4}(a) and  the corresponding R(I) curves are plotted Fig.~\ref{fig4}(c). Experimentally, starting from the $AP$ configuration and above a positive threshold current, the magnetization switches sharply from the $AP$ to $P$ configuration. Then for a larger current, an increase of the resistance is observed due to the onset of a steady precession mode. This behavior is well reproduced by the macrospin simulation (see  Fig.~\ref{fig4}(b)). In particular, the critical currents associated with the magnetization switching from $AP$ to  $P$ state are well reproduced, as well as their dependence with the applied magnetic field.

Now we compare the  experimental (Fig.~\ref{fig4}(c)) and simulated R(I) (Fig.~\ref{fig4}(d)) curves obtained for increasing current starting from an initial $AP$ configuration. A good agreement is obtained (see for example the critical currents or the variation of resistance). In particular, the simulated increase of resistance associated with the transition from the $IP_P$ to the $OP_P$ state is reversible as is the case in experiments. If one calculates the differential resistance dV/dI=R+IdR/dI using the simulated R(I) curves, one observes that  this transition is associated with a peak in the dV/dI(I) curves. This reproduces the experimental observations where such peaks, that characterize the reversibility of the transition, have been always observed   at  the onset of microwave emission (see Ref.~\cite{Boulle07NP}).  In addition, for both simulations and experiments, no significant variation of the resistance is observed for the in-plane precession $IP_P$ mode.  However, the simulation overestimates the stability of the $IP_{AP}$ precession mode around the $AP$ configuration at the expense of the $P$ state (see $IP_{AP}$ zone in Fig.~\ref{fig4}(b) and Fig.~\ref{fig4}(d)).  This can be explained by an underestimation of the effect of temperature caused by a sweeping ramp 10$^5$ faster in simulations than in experiments. With a slower current ramp, the $IP_{AP}$ mode should be more unstable and thus the reversal toward the $P$ state should be favored. This may also explain why we did not measure microwave power associated with this dynamical precession mode.

\subsection{Frequency of the precession}

\begin{figure}[p]
	\centering
		\includegraphics[width=1\textwidth]{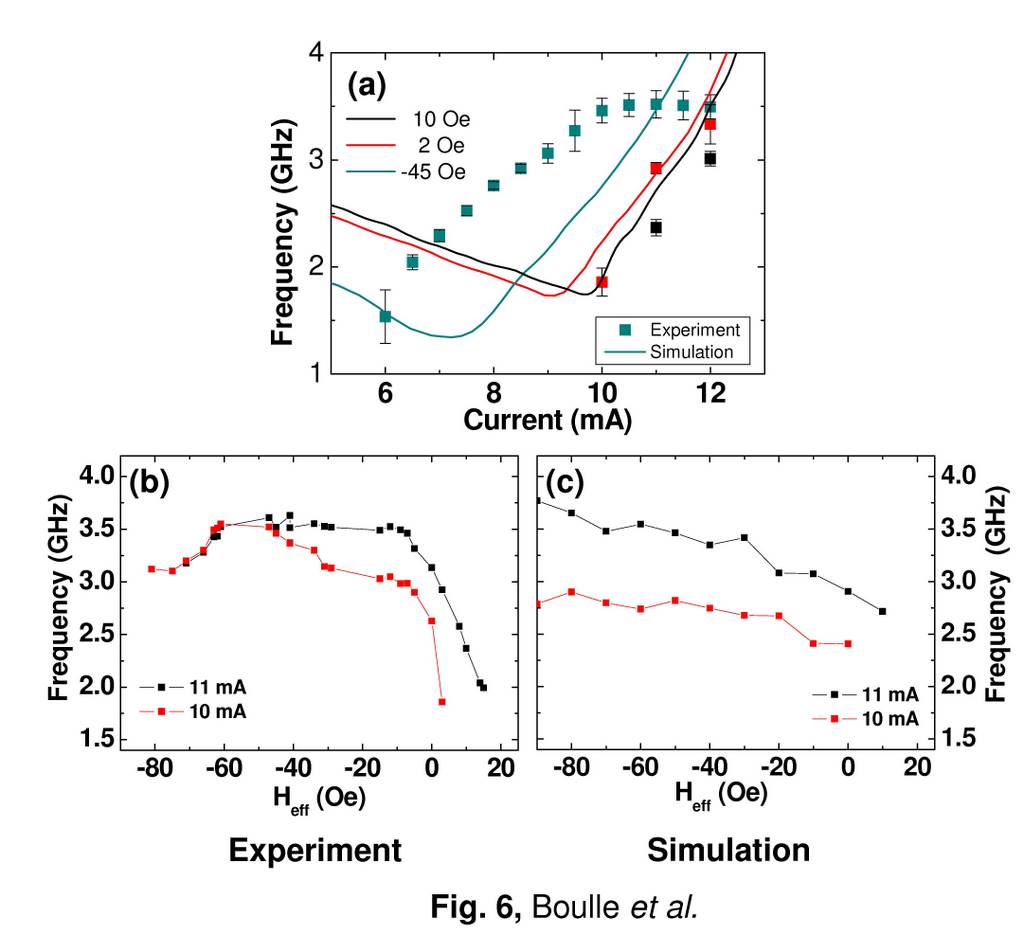}
	\caption{(a) Simulated (plain lines) and experimental (square points) variation of frequency with current for several values of $H_{eff}$. (b-c): Simulated (b) and experimental (c)  variation of frequency with $H_{eff}$ for I=10~mA and I=~11 mA. }
	\label{fig5}
\end{figure}

We compare in Fig.~\ref{fig5}(a) the experimental and calculated f(I) curves for several values of applied field. A good agreement is obtained  for fields close to the zero effective field (H$_{eff}= 2$ and 10~Oe) corresponding to the regime 1 in the phase diagram. In particular, the blue shift with current and the decrease of frequency with field observed experimentally is well  reproduced by the simulation. Simulations and experiments significantly differ  for higher negative field (H$_{eff}=-45$~Oe). In particular, the saturation regime (regime~2) characterized by a frequency approximately constant with current is not reproduced by the simulation.  However, the experimental slope of frequency with current df/dI before saturation (H$_{eff}=-45$~Oe (7~mA$\leqslant I\leqslant9$~mA), H$_{eff}=2$~Oe and H$_{eff}=10$~Oe) is well reproduced by the simulations. This seems to indicate that the regime before saturation (regime~1) is closer  to an homogeneous precession mode than the regimes observed for higher current (regime~2 and 3). 

The disagreement simulation/experiments of the f(I) curve at high negative field reflects a decrease of frequency with field lower in simulations than in experiments. This appears on Fig.~\ref{fig5}(b-c), where the experimental and simulated dependence of the frequency with magnetic field are compared for I=~10 and 11~mA. However, the general shape of the experimental f(H) curve in the red  shift regime for higher field values is qualitatively reproduced by the simulation.

\section{Discussion}
\label{sec:discussion}
The macrospin simulations well reproduce in overall the main experimental features: the region in current-field diagram with larger power emission and increased resistance, the frontiers between the main different static and dynamical states, and the dc resistance of these different states. This agreement  is quite satisfactory if we consider the crude macrospin approximation that was used. 

The direct comparison of the experimental and simulated dynamical phase diagrams clearly suggests that the observed excitations are associated with an out-of-plane precession mode. This is confirmed by the  blue shift in current  and the red shift in field observed in the dynamical regime 1, signature of this dynamical  mode.
However,  the macrospin simulation  do not reproduce the saturation regimes observed   at  low applied field and for a large current (regime~2 and 3) (see Fig.~\ref{fig5}(a)). In this range of parameters, we believe that these regimes are probably associated to an inhomogeneous distribution of magnetization. It is worth noting that the only two other groups who observed a blue shift regime for an in-plane applied field, did also observe this saturation (or decrease)  of the frequency for larger applied current (see Fig.~3, Ref.~\cite{Kiselev05PRB} and Ref.~\cite{Houssameddine07NM}), suggesting the intrinsic character of this dynamical feature. On the basis of micromagnetic simulations~\cite{Houssameddine07NM}, it has been recently suggested that this behavior can be attributed to the onset of inhomogeneous dynamical modes characterized by a mean out-of-plane component of magnetization $<m_z>$ lower than for the uniform mode. The onset of such non standard modes leads to a decrease of the large magnetostatic energy generated by the strong demagnetizing field when the magnetization goes out of the film plane. The frequency of  precession being proportional to $<m_z>$, the decrease of $<m_z>$ results in a decrease of the frequency as compared to the macrospin prediction and thus explains this saturation phenomenon. The in-plane Oersted field may also play a role: in the case of an out-of-plane precession mode, it tends to pull the magnetization towards the film plane. This also leads to a decrease of the precession frequency~\cite{Loubens05}.

Another point to discuss is, as already mentioned, the non-observation of a red shift regime associated with the $IP_P$ mode in our experiments. This regime was  not observed experimentally in the Co(8~nm)/Cu/Py(8~nm) nanopillar described in this paper.  However, we could observe it in a nanopillar having a different WAD-STT structure Co(4~nm)/Cu/Py(8~nm). In this nanopillar, before the onset of the higher power blue shifting microwave peaks,  signals of very low amplitude presenting a clear red shift behaviour have been observed. In \textit{standard} nanopillars, this $IP_P$ mode corresponding to a large precession angle goes with a large emitted power~\cite{Kiselev03N,Sankey05PRB}. This difference between \textit{standard} and WAD-STT samples can be explained by a weaker slope of the angular dependence of the GMR for angles around $\varphi=0$ in the case of a WAD-STT ~\cite{Urazhdin05PRB}. This results in a weak variation of the resistance (and therefore a very weak power) for in-plane oscillations of magnetization even for large angles. In our calculations, we accounted for  this specific angular dependence of the GMR by using  a normalized resistance $r(\varphi)=\sin^2(\varphi/2)/(1+\chi\cos^2(\varphi/2))$ with an asymmetry factor~\cite{Urazhdin05PRB}  $\chi=7.7$. As can be seen on Fig.~\ref{fig2}(b) or~ref{fig4}(d)  , this results in a low increase of resistance of  the device in the $IP_P$ mode. To evaluate the influence of the angular dependence of the GMR on the emitted power, we calculated the maximum emitted power for this $IP_P$ mode using Eq.~\ref{eq:ang} with $\chi=7.7$, $\chi=2$ (measured experimentally by Urazhdin \emph{et al.} in a \textit{standard} structure Py(6~nm)/Cu/Py(12~nm), and $\chi=0$) corresponding to the simple angular dependence $\Delta R=(1-\cos\theta)/2$. The simulated power in the precession mode for $\chi=7.7$ is about 10 times lower than for $\chi=2$ and 25 times lower than for $\chi=0$. If this assumption is valid, then it allows to explain that in our experimental configuration, the measured signals in this $IP_P$ precession mode were too weak to be detected.

Finally, our simulations allow the calculation of the output power in the $OP_P$ mode: it is about 3.9 pW/mA$^2$, should compare to the maximum measured output power of about 4.5x10$^-2$ pW/mA$^2$, \emph{i.e} approximately a factor 80 smaller. This discrepancy may arise from several factors. First, due to the impedance mismatch and the attenuation of the signal in the high frequency line and the variation of the amplification  gain with frequency, the measured signal (after taking into account the amplification) is only a fraction of the actual emitted signal. Second, the fact that the $OP_P$ mode predicted by the macrospin simulation has been rarely observed experimentally~\cite{Kiselev03N} indicates the macrospin approximation does not describe properly the dynamics in this precession mode. Recently, micromagnetic calculations by Berkov~\emph{et al.}~\cite{Berkov05PRBa, Krivorotov07PRB} have shown  that this dynamical $OP_P$ mode is sensitive to the magnetization homogeneity. Factors favoring the inhomogeneity such as temperature or the Oersted field decrease the output power in this precession mode. Third, the possible presence of ferromagnetic oxydes (NiO) on the sidewalls of the pillar due to air exposure during the fabrication process~\cite{Emley06PRL} or the stronger Oersted field on the edges may force the magnetization to lie in the plane on the edges. In this case, only a part of the magnetization in the center might therefore precess in the out-of-plane direction, the resulting output power being weak.

\section{Conclusion}
As a conclusion, we have simulated the magnetization dynamics in the macrospin approximation taking into account the WAD-STT predicted for a Co(8~nm)/Cu/Py(8~nm) nanopillar by the Barna\'s-Fert model~\cite{Barnas05PRB,Barnas06MSaEB,Boulle07NP}. At low field, the magnetization dynamics is strongly modified as compared to a standard angular dependence of the torque.  When the current is increased, at low and even zero applied field, one observes successively an in-plane precession mode ($IP_P$) in which frequency decreases with current ("red shift regime") and an out-of-plane precession mode ($OP_P$), in which the frequency increases with current ("blue shift regime"). The transition between both regimes is associated with an  increase of  the dc resistance and a strong increase of the output power. We have presented  some experimental results that confirm this zero and low field steady precession. In the main part of the current-field phase diagram in the low field range, the frequency increases with current (blue shift regime) and decreases with field. In the remaining part of the diagram (large negative applied field and high applied current), a regime in which the frequency is approximately constant (or decreases slightly) with current is observed. The macrospin simulations reproduce the main observed experimental  features  (phase diagram, R(I) curves, dependence of frequency with current and field) and suggest that the dynamical excitations observed experimentally are associated with an out-of-plane precession mode. However the saturation regimes observed at lower field and higher current are not reproduced by the simulation and may reveal the onset of inhomogeneous dynamical excitations. Following the first experimental confirmation~\cite{Boulle07NP} of the STT induced zero  field steady precession  caused by the WAD-STT, this detailed comparison of the static and dynamical results between models and experiments and the observed general  good agreement give an additional confirmation of the predicted WAD-STT by diffusive models of spin transfer. This possibility of engineering the angular dependence of the torque with different spin diffusion lengths in the magnetic layers underlines the role of the spin accumulation and the spin relaxation effects on the STT. By playing on the distribution of spin relaxation in the structure, it is therefore possible to strongly modify the STT and the magnetization dynamics  induced by the torque.

\section{Acknowledgements}
We would like  to acknowledge L. Vila for assistance in fabrication,  H. Hurdequint for FMR measurements, O. Copie and B. Marcilhac for assistance in transport and frequency measurements. This work was partly supported by the French National Agency of Research ANR through the PNANO program (MAGICO PNANO-05-044-02) and the EU through the Marie Curie Training network SPINSWITCH (MRTN-CT-2006-035327). 
\appendix
\section{Parameters used for the macrospin simulation}
\label{appendix1}
For the macrospin simulation a gyromagnetic ratio $\gamma_0=2.2.10^{5}$~m/(A.s) has been used. The saturation magnetization $\mu_0M_S=0,87$~T has been  deduced from ferromagnetic resonance experiments carried out on Cu(6nm)/Py(7nm)/Cu(6nm) thin films.  The  anisotropy fields  $H_d$ and $H_{an}$ can be expressed as a function of demagnetizing factor $N_x$, $N_y$ et $N_z$ associated with the shape of the nanomagnet with $H_d\approx M_s(N_z-N_y)$ et $H_{an}\approx M_s(N_y-N_x)$. If we approximate the shape of the nanomagnet with  an ellipsoid  with axes 150~nm, 105~nm and 8~nm, close to the lateral dimensions measured by SEM within measurement uncertainty, one obtains~\cite{Osborn45PR} Nx=0.047, Ny=0.063, Nz=0.89  leading to $\mu_0H_d\approx0.7$~T and $\mu_0H_{an}\approx0.014$~T. This value of $H_{an}$ is close to the anisotropy field $\mu_0H'_{an}\approx0.0145$~T one can estimate from the room temperature experimental coercive field $H_c$\,\footnote{The coercive field $H_c$ at a temperature T can be expressed as a function of the uniaxial anisotropy field  $H'_{an}$~\cite{Sharrock90IToM}  $H_c(H'_{an})=H_{an}(1-[k_BT/E_k\ln(f_0\tau_m/\ln2)]^{1/2})$ with  $E_k=1/2\mu_0M_sVH'_{an}$, $f_0$ the attempt frequency of the order of 1~GHz and $\tau_m$ the time measurement of the order of 1s. By solving this equation, using experimental parameters and assuming $f_0$=1~GHz, one finds $H'_{an}\approx145$~Oe.}. For a direct comparaison between experimental and theoretical phase diagrams, the uniaxial anisotropy field $H_{an}$ was adjusted to obtain a simulated coercive field equal to the experimental one ($\mu_0 H_c=0.009$~T). For simulation at T=0~K, we used therefore $\mu_0 H_{an}=0.009$~T.   At T=300~K and with the parameters and waiting time used in the simulation, one obtains $\mu_0 H_{an}=0.012$~T. A Gilbert damping parameter $\alpha=0.018$ has been used. Finally, currents are deduced from current densities using a lateral surface of the nanopillar of A=1.38x10$^{-14}$ m$^2$, measured experimentally by SEM.
%The Gilbert damping parameter $\alpha$  has been chosen so that the critical currents for excitations calculated in the 300~K simulations are closed to the experimental ones (currents are deduced from current densities using a lateral surface of the nanopillar of A=1.38x10$^4$ nm$^2$, measured experimentally by SEM). Using this procedure, one finds $\alpha=0,018$ in the range of   experimental  values reported in Py nanopillar ($0.01\leqslant\alpha\leqslant0.035$)~\cite{Krivorotov05S,Braganca05APL,Emley06PRL,Fuchs07APL}.
Since the angular dependence of the GMR in a  WAD-STT structure is predicted~\cite{Manschot04PRB,Urazhdin05PRB} to deviate significantly from the commonly used dependence   $\sin^2(\varphi/2)$, the normalized magnetoresistance $r= (R-R_P)/(R_{AP}-R_{P})$ was calculated using the following relation~\cite{Slonczewski02JoMaMM,Manchon06PRBa}:
\begin{equation}
\label{eq:ang}
	r(\varphi)=\sin^2(\varphi/2)/(1+\chi\cos^2(\varphi/2))
\end{equation}
  We have used $\chi=7.7$ that was derived experimentally by 
Urazhdin~\emph{et al.}~\cite{Urazhdin05PRB} in a non symmetrical pillar composed of Py(6~nm)/Cu(10~nm)/Py(1.5~nm) for which a WAD-STT is also predicted.

The  precession frequency   was derived from the higher amplitude  lowest order peak in the Fourier transform spectrum of the $m_y$ component. In the  case, very common experimentally, of a slight misalignment between the magnetization of the polarizing layer and the easy axis of the free layer, this frequency corresponds actually to the higher amplitude lowest order frequency measured experimentally~\cite{Montigny05JoAP}\,\footnote{ One can note however that the frequency derived from the m$_x$ and the m$_y$ component are the same for  an out-of-plane precession mode.}.
The output power P corresponding to the simulated trajectories is deduced from the reduced resistance r using  $P/I^2=\Delta R_{exp}^2/Z_c.<(r-<r>)^2>$, with $\Delta R_{exp}=R_{AP}-R_P=51~m\Omega$ the experimental static variation of resistance due to the GMR, and $Z_c=50~\Omega$ the characteristic impedance of the high frequency line.
$<...>$ indicates the mean value of 40~ns after relaxation of the magnetization.
%La puissance hyperfréquence   correspondant aux  trajectoires simulées est déduite de la résistance réduite r  suivant la relation $P/I^2=\Delta R_{exp}^2/Z_c.<(r-<r>)^2>$, avec $\Delta R_{exp}=R_{AP}-R_P=51~m\Omega$ la variation de résistance statique expérimentale due à la GMR  et $Z_c$, l'impédance caractéristique  de la ligne égale à 50~$\Omega$. $<...>$ indique la valeur moyenne sur 40~ns après relaxation de l'aimantation. 

 The LLG equation~(\ref{eq:LLG}) has been solved using a fourth order Runger-Kutta algorithm with a time step of 1~ps. For simulation at T=0~K, to let  magnetization reach a stationnary state, a 100 ns relaxation time has been used, then trajectories are saved over a 40~ns. These trajectories are used for the fast Fourier transform calculations and the deduction of the magnetoresistance.  For simulation at T=300~K, after application of the current, a relaxation time of 30~ns has been used and simulations have been carried out by sweeping current with a 10$^5$ mA/s  ramp. The plotted frequency and resistance vs current curves are averaged over 5 realizations or more.

\end{document}